\documentstyle[aasms4,12pt]{article}
\begin{document}

\title{Space Telescope Imaging Spectrograph Parallel Observations 
   of the Planetary Nebula M94-20
\footnote{Based on observations made with the NASA/ESA
Hubble Space Telescope, obtained from the data archive at the Space 
Telescope Science Institute. STScI is operated by the Association 
of Universities for Research in Astronomy, Inc., under the NASA 
contract NAS 5-26555.}}

\author{Philip Plait\altaffilmark{2,3} and Theodore R. Gull\altaffilmark{3}}

\affil{$^2$Advanced Computer Concepts, Inc., Potomac, MD 20854 USA;
plait@abba.gsfc.nasa.gov}
\authoremail{plait@abba.gsfc.nasa.gov}

\affil{$^3$Laboratory for Astronomy and Solar Physics, Code 681,
Goddard Space Flight Center, Greenbelt, MD 20771 USA; gull@sea.gsfc.nasa.gov}
\authoremail{gull@sea.gsfc.nasa.gov}

\begin{abstract}

The planetary nebula M94--20 in the Large Magellanic Cloud
was serendipitously observed with the Space Telescope Imaging
Spectrograph on board the Hubble Space Telescope as part of the
Hubble Space Telescope Archival Pure Parallel Program. 
We present spatially resolved imaging and spectral data of
the nebula and compare them with ground based data, including 
detection of several emission lines from the nebula and
the detection of the central star.
We find the total 
${\rm H\alpha + [NII] \ flux  = 7.3 x 10^{-15} \ erg \ s^{-1} \ cm^{-2} }$
and we estimate the magnitude of the central star to 
be $\rm{m_{V}} = 26.0 \pm 0.2$. Many other ${\rm H\alpha}$ sources
have been found in M31, M33 and NGC 205 as well.
We discuss the use of the parallel
observations as a versatile tool for planetary nebula surveys 
and for other fields of astronomical research.

\end{abstract}


\section{Introduction}

The Hubble Space Telescope (HST) Archival Pure Parallel Program,
started in 1997 shortly after the second Hubble servicing mission,
was implemented 
to provide the maximum amount of science possible
with HST and its primary cameras:
the Wide--Field Planetary Camera 2 (WFPC2), the Near-Infrared 
Camera and Multi--Object Spectrograph (NICMOS), and the
Space Telescope Imaging Spectrograph (STIS). During observations of a
target by the primary instrument, the other instruments make a 
preplanned series of observations in nearby fields,
the positions of which are determined by the focal plane
offset of the instrument and the roll angle of HST. The resulting
observations are placed in the Hubble Data Archive
at the Space Telescope Science Institute, and are made immediately
available to the astronomical community.

The STIS parallels include broadband camera images and full field
slitless spectroscopy. Full details
of the STIS parallel survey (SPS) capabilities and techniques can be 
found in Gardner et al. (1998). 
The imaging mode has a 
very broad bandpass (FWHM $\approx$ 5000\AA)
peaking at 5500\AA\ and ranging from $\approx$ 2500\AA\ to 10300\AA. 
The full field
spectroscopy mode has a wavelength range of 5200\AA\ to
10300\AA, and a dispersion of 4.9\AA\ per pixel.
In an imaging mode observation of 2000 seconds, point sources 
as faint as ${\rm m_V}=28$ can be detected with a S/N of 5,
and the spectra can achieve a limiting magnitude of 22 under
the same circumstances. 

Approximately 6000 STIS parallel images and spectra were taken
in the first year of the parallel program, comprising more than 
1000 separate fields. Of these,
over 300 images and spectra are of the Large Magellanic Cloud (LMC). 
These data represent a unique opportunity to peer deeply into the LMC
and investigate many of its physical properties. 
Planetary nebulae (PNe, singular PN) provide a way to study several 
important properties of the LMC, including kinematics (Vassiliadis,
Meatheringham \& Dopita 1992), dynamical and chemical evolution
and can also be used as distance indicators (for example, Feldmeier,
Ciardullo \& Jacoby 1997 and references therein).
Nearly 300 PNe have been found in the LMC
(Leisy et al. 1997).
LMC PNe are of particular importance in the determination of the
PNe luminosity function, since they all lie at approximately the
same distance. Unfortunately, the great distance to the LMC means
that most PNe are very faint and unresolvable by ground based
telescopes, and the typically crowded LMC fields make
studying the PNe very difficult.

The high spatial resolution of HST and low sky background 
of Earth orbit
provide an advantage for studying LMC PNe
(for example, Dopita et al. 1996).
In particular, STIS can record spatially resolved spectra of
the larger, fainter PNe. Here we report on the
serendipitous STIS parallel observations of the 
planetary nebula M94--20, first discovered by Morgan (1994) in a
ground--based emission line survey of the LMC. The STIS camera mode 
images clearly resolve the nebula to be $\approx$2 arcseconds in
diameter and also detect the central star. The
spectra provide identification of several emission
lines in the nebula. The low resolution of the STIS spectra 
together with the relatively large angular extent of the nebula blends
the nebular diagnostic emission lines, but
the images and spectra are still a useful 
tool in investigating extragalactic PNe and can be used to plan
followup observations, both ground and space--based.

\section{Observations}

Six images and two spectra were taken of the field
containing the planetary nebula M94-20. The total 
exposure times were 1100 and 1200 
seconds for the images and spectra, respectively.
The observations were taken as part of the STIS 
Archival Pure Parallel Program, HST Program ID 7783, 
S. Baum, Principal Investigator.
WFPC2 was the prime instrument during the STIS parallels, 
observing the LMC as part of the
HST program ``Star Formation History
of the Large Magellanic Cloud'', HST ID 7382, 
Smecker--Hane, Principal Investigator.
The primary target was
LMC-DISK1 at $\alpha = 05^h 11^m 14\farcs92$ \ and \ $
\delta =-71^{\circ} 15^m 41\farcs62$. STIS was
located approximately five arcminutes north of WFPC2 during the 
observations. The STIS parallel images were formatted as 1024x1024 
arrays (one pixel = 0\farcs 051), while the spectra were automatically
binned onchip in the y--direction to 1024x512 (one pixel 
= 0\farcs 051 in the spectral direction, and 0\farcs 102 in the 
spatial direction).
The observations were taken over two orbits, and
there was a 0\farcs8 (15.5 STIS CCD pixels) telescope slew 
to the southwest between the two sets of 
observations. The offsets between images were determined by 
cross--correlating a small subsection in each, which were then
used to shift and combine the images. The offsets were applied with
appropriate binning to the spectra which were shifted and combined
as well. The images and full field spectra were processed
using STIS Investigation Definition Team pipeline calibration
software which performs basic data reduction steps such as bias
and dark current subtraction. The observation particulars 
are shown in Table 1.

For the analysis using the imaging mode, the sky background was
subtracted by taking a simple median of the area near the PN. 
The background in the full field spectral image
was subtracted from the spectrum using a column--by--column
median, employing sigma--clipping to remove the 
positive bias of the stellar contamination.

\section{Discussion}
\subsection{The Nebula}

Figure 1 shows the full field processed image containing M94--20.
The PN is located at the top of the image, centered roughly in the 
horizontal (x--axis) direction. 
North is to the left and East is down; we display the image
in this way so that it has the same sense as 
the spectrum, which disperses light along the x--direction.
Though close to the detector edge, the
nebula is located fully inside the image.
Note the small irregular galaxy located 9 arcseconds
to the north of M94--20, and another located 22 arcseconds to the south.
Although both lie in the dispersion direction of the PN, neither 
interferes significantly with the spectrum.
The bright diagonal line in the image
is a diffraction spike from a star located just off the
detector, and the circular features near bright stars are internal
reflections in STIS. The inset shows a close--up of just the nebula.
The PN, unresolved in the discovery survey (Morgan 1994),
is clearly resolved in the STIS image. 
The measured properties of the nebula and central star (see Section 3.2)
are listed in Table 2. The nebula is slightly elliptical, with a
mean diameter of roughly 2 arcseconds, making this one
of the largest PNe in the LMC. Most LMC PNe are smaller than 1 arcsecond
in diameter, with the notable exception of LMC--SMP72,
a bipolar PN measuring approximately 2 x 3 arcseconds 
(Dopita et al. 1996). M94--20 has a bright
elliptical rim, and the 
outermost parts of the nebula also show
some faint structure, reminiscent of the double--elliptical
structure of NGC 6543 (aka the ``Cat's Eye''). One of the outer
ellipses is aligned with the inner rim, while the other
has a position angle of $\sim 145^{\circ}$.
The large angular extent of the PN implies that this
is an evolved object.

A subarray of the spectrum containing M94--20 is shown in Figure 2
({\it top}). The extracted subarray covers the full spectral range
of the original spectrum, but only 32 pixels ($3\farcs3$) 
in the spatial direction. The stars
in the image appear as sources of continuum, while M94--20
is clearly an emission--line object. 
Figure 2 ({\it middle}) shows a closeup of the nebular
spectrum from 6000\AA\ to 7200\AA\ as well as
the positions of the line images
using ellipses with major and minor axes corresponding to
those measured in the camera mode image ({\it bottom}).
The bright elliptical patch 
corresponds to the lines of H$\alpha$ and [NII] in M94--20.
At this resolution, the [NII] 6548\AA\ and 6584\AA\
lines are separated from H$\alpha$ by only 3 and 4 pixels,
respectively. The nebula itself is about 40 pixels or ten times
that size, so the three spectral line images overlap.
The blended nebular spectral image is
about 4 pixels wider than the nebular image in the camera
mode, also indicating that more than one line is present.
The [OI] line at 6300\AA\ is also clearly seen.
The [SII] 6717, 6731\AA\ emission
line images are barely detected in the spectrum. 
We also note that we may have a detection of
[SIII] 9069\AA\ and Pa$\epsilon$,[SIII] at 9545\AA, which
does not reproduce well in Figure 2 but can be seen very faintly in
the original data.

We flux calibrated the brighter emission lines using the 
absolute sensitivity of STIS for a point source
(Collins and Bohlin 1998), correcting the sensitivity curve for a
slitless spectrum of an extended line emission object, 
and masking out the continuum spectra of nearby stars that overlapped the
PN spectrum.
We find that the 
total observed H$\alpha$ + [NII] 6548, 6584\AA\ flux is 
$\rm{7.3 x 10^{-15}\ erg/sec/cm^2}$.
D. Morgan and Q. Parker (1998, private communication) 
made followup observations of many of the 
PNe found by Morgan (1994),
and for M94--20 detected [OIII] 4959, 5007\AA\
and H$\alpha$. They report the [OIII] flux to 
be $\rm{2.7 x 10^{-14}\ erg/sec/cm^2}$. 
The value of
the ratio of H$\alpha$ + [NII] 6548, 6584\AA\ to [OIII] 4959, 5007\AA\ 
in LMC PNe is typically
3--4 (for example, Vassiliadis, Dopita, Morgan and Bell 1992);
which is somewhat lower than but consistent with these measurements.
The typical value of the ratio of [OI] 6300\AA\ to H$\alpha$
in LMC PNe is approximately 0.07. We find the [OI] 6300\AA\ flux is
$\rm{1.2 x 10^{-15}\ erg/sec/cm^2}$, which again gives a ratio higher
than but consistent with the typical value.  Morgan and Parker
also find an upper limit to the H$\beta$ flux of
$\rm{1.2 x 10^{-14}\ erg/sec/cm^2}$, making
M94--20 a relatively faint LMC PN (Vassiliadis, Dopita, 
Morgan and Bell 1992).
We note that our measurements may also be
upper limits, since background subtraction is difficult due
to the stellar spectra superposed on the PN spectrum.
The detection of [SII] 6717, 6731\AA\ is very weak
and is complicated by their proximity to the 
H$\alpha$ + [NII] 6548, 6584\AA\ lines. We subtracted a linear fit 
to the slope of the wing of the H$\alpha$ + [NII] lines to find the total
flux of the [SII] lines, and get an upper limit of 
$\rm{5 x 10^{-16}\ erg/sec/cm^2}$.

The emission lines from the PN are remarkably featureless spatially, 
given the obvious structure in the camera mode image. The 
spectral mode of STIS has a blue cutoff at $\sim$5300\AA, while the 
imaging mode goes down to $\sim$2000\AA. The imaging mode
will therefore detect the lines of H$\beta$,
the [OII] doublet at 3727 and 3729\AA, and the lines of 
[OIII] at 4959, 5007\AA. The emission lines seen in the imaging mode but not 
in the spectroscopic mode must account for these features. 

\subsection{The Central Star}

The central star, as well as two other superposed stars, are clearly
visible in the camera mode image.
Relative to the Fine Guidance Sensor guide stars, we find the 
J2000 coordinates of the central star are 
$\rm{\alpha = 5^h 11^m 10\farcs 64 \ and \
\delta = -71^{\circ} 10^m 26\farcs 54}$, in good agreement with
those found in Leisy et al. (1997). The spectrum of the
star is too weak to determine a temperature, 
so we used the camera mode sensitivity of STIS to flux calibrate
blackbody models of 20000K, 40000K and 100000K. The calibrated
models were then folded into a Johnson V filter bandpass
to find the V magnitude of the star. 
We found the central star magnitude to be $\rm{m_{V}} = 26.0 \pm 0.2$.
The sensitivity of STIS in the imaging mode drops rapidly
with decreasing wavelength, so the calculated magnitude does not
depend strongly on the temperature of the star. 
Given the magnitude and a distance to the LMC of 55 kpc, we find that
the luminosity of the central star is 0.09 solar, and 
${\rm log(T_{eff}/T_{\odot}) = 4.6}$ for a typical white dwarf radius
of ${\rm 5 x 10^8 cm}$. The low temperature indicates this
may be a relatively old white dwarf, which is consistent with the
evolved nature of the PN.

\subsection{Other PNe}

There are approximately 300 known PNe in the LMC 
(Leisy et al. 1997). The number density is highest
near the Bar and tapers off rapidly with distance
(Morgan 1994). M94--20 lies  over a degree from the Bar,
near the edge of the LMC. Over 300
observations comprising 70 separate fields have
been taken in the LMC by STIS. Of these, 29 also 
have associated full field spectra, enhancing the 
ability to detect candidate PNe (i.e., emission 
line objects) clearly. 
The detection limit of an observation depends most
strongly on the brightness and the size of the nebula in a given
emission line and the integration time. The 1200 second
exposure of the M94--20 observations is fairly typical of
parallels, and noting that the [SII] 6717, 6731\AA\ emission
lines in the nebula are barely detected, we expect to detect
any nebula the size of M94--20 brighter than about 
${\rm 2-3 x 10^{-15}\ erg/sec/cm^2}$ in a line for a typical LMC PN.
A smaller nebula would be detectable at fainter limits,
with the brightness scaling inversely with the area.
We note here again that M94--20 is a relatively faint, large LMC PN.

No other previously catalogued LMC PNe have been observed by
STIS in the parallel survey.
We have processed and examined the other fields in the LMC and
also 20 STIS fields in the Small Magellanic Cloud (SMC)
and have found no other extended PNe 
to within the detection limits, catalogued or otherwise, 
although large HII regions are common
and easily detected. Unresolved or partially
resolved PNe are difficult to identify in this case because most
of the full--field spectra in the LMC and SMC are either single observations
or have one cosmic ray split (that is, two observations without
telescope movement), making the differentiation of sharp
spectral features from cosmic rays and hot pixels
difficult. We note here that in many other fields, STIS
observations have two or more cosmic ray splits, making unresolved PN
detection relatively easier. We have also searched the
relevant WFPC2 parallel fields;
no known LMC or SMC PNe are located within these parallel 
survey fields.
Discovering previously unrecorded PNe using WFPC2
parallels is more difficult due to the lack of 
unambiguous spectral information in the broad 
passbands used.

Although the odds of finding PNe in the random STIS parallels
are low for the LMC or SMC, we are encouraged by the observations 
of M94--20. The ability of STIS to get deep, spatially resolved
spectra makes it an excellent instrument for tagging objects for
further deep ground--based spectroscopy.
STIS parallels will continue to be taken in the 
Magellanic Clouds, and it is only a matter of time before additional
PNe are observed. Interestingly, PNe in nearby galaxies can also
be detected in STIS parallels, although this becomes a difficult
process as PNe at that distance are no longer resolved spatially.
However, it also means that the larger volumes of space on 
the scale of the host galaxy are observed as well.
Ground--based surveys using narrow passband filters 
centered on [OIII] 5007\AA\ and H$\alpha$
have found many PN candidates in M31 and M33
(Ciardullo et al. 1989; Bohannan, Conti \&  Massey 1985).
We made a preliminary search of STIS parallel fields in M31 and NGC 205
which yielded several emission lines objects. At least one of these 
objects
is a planetary nebula, even though only a 
small fraction of the surface area of those galaxies has been observed.
A similar search in M33 fields has also resulted in several detections.
These observations are awaiting ground--based followups to get
spectra with wavelength coverage that includes ${\rm H\beta}$
and [OIII] 5007\AA\ to help distinguish PNe from HII regions.
Future STIS parallels will undoubtedly find more PNe,
especially if the narrow [OII] 3727\AA\ and
[OIII] 5007\AA\ filters are used in imaging mode in conjunction
with the G430L grating, which has a bandpass that includes these
emission lines.

We note finally that the use of STIS parallels goes well
beyond that of finding (or adding to our knowledge of pre--existing)
PNe. Moderate redshift galaxy
counts (Gardner et al. 1998), research on 
stellar population counts, low mass stars (Plait 1999),
globular clusters in nearby galaxies, 
and many other fields can benefit from a 
relatively simple search of the parallel archive. 
Since the data already exist and are publically accessible, 
the STIS parallels are a excellent tool that should be exploited to 
their fullest extent.

The authors wish to acknowledge the help of 
Nick Collins, Bob Hill, Jon Gardner and David Morgan
for giving their advice during this work.

\newpage

Figure 1: Combined and processed camera--mode image of M94--20
field. The inset shows a close--up of the nebula. The inner 
rim and central star can be seen clearly in the close--up image.

\bigskip
\bigskip

Figure 2.-- ({\it Top}) A subarray of the full field spectrum
containing M94--20. The subarray covers the full spectral range of the
original spectrum, and 32 pixels ($3\farcs3$) in the spatial
direction. Several emission lines can be seen in
this log scale image, while stars appear as streaks. 
The close--up of the region from 6000\AA\ to 7200\AA\ 
is also shown ({\it middle}) along with a schematic of the positions of
the detected emission lines ({\it bottom}).

\newpage

\begin{tabular}{lcccc}
\multicolumn{5}{c}{Table 1}\\
\multicolumn{5}{c}{Log of STIS M94--20 Observations (1997 October 23)}\\ 
\hline \hline
Rootname & STIS Mode & Exposure &
   R.A. & Dec \\ 
& & (sec) & (2000) & (2000) \\ 
\hline
O48B74010 & IMAGE  & 400 & 5 11 15.44 & -71 10 30.9 \\
O48B75010 & IMAGE  & 400 & 5 11 15.56 & -71 10 30.3 \\
O46N43FGQ & IMAGE  & 150 & 5 11 15.44 & -71 10 30.9 \\
O46N43FHQ & SPEC   & 600 & 5 11 15.44 & -71 10 30.9 \\
O46N44FRQ & IMAGE  & 150 & 5 11 15.56 & -71 10 30.3 \\
O46N44FUQ & SPEC   & 600 & 5 11 15.56 & -71 10 30.3 \\
\hline
\end{tabular}

\newpage

\begin{tabular}{ll}
\multicolumn{2}{c} {Table 2} \\
\multicolumn{2}{c} {Properties of M94--20 and the central star}\\ 
\hline
\hline
Parameter & Value\\
\hline
\multicolumn{2}{c} {Nebula}\\
\hline
Major axis ................. & $2\farcs1$  \\
Minor axis ................. & $1\farcs6$ \\
Major axis (rim) ........ & $1\farcs3$ \\
Minor axis (rim) ........ & $0\farcs9$ \\
P.A. (rim) .................. & $30^{\circ}$ \\
H$\alpha$ + [NII] flux ..........& ${\rm 7.3 x 10^{-15} \ erg \ s^{-1} \ cm^{-2} }$ \\
{}[OI] $\lambda$6300 flux ...........& ${\rm 1.2 x 10^{-15} \ erg \ s^{-1} \ cm^{-2} }$\\

\hline

\multicolumn{2}{c} {Central Star} \\

\hline

${\rm m_v}$ ..............................& 26 $\pm$ 0.2 \\
${\rm log(L/L_{\odot}) }$.................... & $-1.0$ \\
${\rm log(T_{eff}) }$ ......................& 4.6 \\

\hline
\end{tabular}
\bigskip

\end{document}